\documentclass[letterpaper,12pt]{JHEP3}
\usepackage{bm}
\usepackage{amsmath}
\usepackage{yfonts}
\newcommand{\qn}{\textswab{q}}
\newcommand{\wn}{\textswab{w}}
\newcommand{\<}{\langle}
\renewcommand{\>}{\rangle}
\renewcommand{\d}{\partial}
\newcommand{\N}{{\cal N}}
\renewcommand{\O}{\hat{\cal O}}
\newcommand{\q}{\bm{q}}
\newcommand{\x}{\bm{x}}

\newcommand{\gYM}{g_{\mathrm{YM}}}

\newcommand{\stru}{\rule[-.2in]{0in}{.2in}}

\title{From AdS/CFT correspondence to hydrodynamics}

\author{Giuseppe Policastro\\
Scuola Normale Superiore, Piazza dei Cavalieri 7, 56100,
Pisa, Italy\\
Email: \email{g.policastro@sns.it}
}
\author{Dam T.~Son and Andrei O.~Starinets\\
Institute for Nuclear Theory, University of Washington,
Seattle, WA 98195, USA\\
Emails: \email{son@phys.washington.edu, starina@phys.washington.edu}
}

\preprint{INT-PUB 02-32}

\date{April 2002}

\abstract{
We compute the correlation functions of R-charge currents and
components of the stress-energy tensor in the strongly coupled
large-$N$ finite-temperature $\N=4$ supersymmetric Yang-Mills theory,
following a recently formulated Minkowskian AdS/CFT prescription.  We
observe that in the long-distance, low-frequency limit, such
correlators have the form dictated by hydrodynamics.  We deduce from
the calculations the R-charge diffusion constant and the shear
viscosity.  The value for the latter is in agreement with an earlier
calculation based on the Kubo formula and absorption by black branes.
}

\begin{document}
\section{Introduction}

Recently, the correspondence between conformal field theories (CFT)
and string theories or supergravity on certain background has been
under intense investigation~\cite{Maldacena,GKP,Witten1,Aharony:1999ti}.
The
original and best studied example is the correspondence between the
$\N=4$ supersymmetric Yang-Mills (SYM) theory in the large $N$ limit
and the type IIB string theory in $\mathrm{AdS}_5\times \mathrm{S^5}$ 
space.  In the
strong coupling limit of the field theory, the string theory is
reduced to classical supergravity, which allows one to calculate all
field-theory correlation functions.

One of the first nontrivial predictions of the gauge theory/gravity
correspondence was that for the entropy of the $\N=4$ SYM
theory at finite temperature~\cite{Gubser:1996de}.  According to
gravity calculations, the entropy of this theory in the limit of large
't Hooft coupling is precisely 3/4 times the value in the zero
coupling limit.  However, so far there has been no independent check
of this result from the field-theoretical side. Indeed, due to the lack of 
supersymmetry at finite temperature no non-renormalization theorem is expected
to hold, while the strong coupling calculations are notoriously difficult.

A serious effort has been recently invested in an attempt to build a
quantitatively predictive gravity dual for a realistic gauge theory.
With this work as yet incomplete,
one nevertheless has discovered many aspects resembling those in the
theory of strong interactions within the framework of the AdS/CFT
correspondence~\cite{Polchinski:2001tt,Giddings:2002cd,Gubser:2002tv}.

In this paper, we use the hydrodynamic limit as a 
nontrivial
test of the AdS/CFT correspondence at finite temperature.  The basic
idea is very simple: the long-distance, low-frequency behavior of any
interacting theory at finite temperature must be described by the
old-fashioned theory of fluid mechanics 
(hydrodynamics)~\cite{Landafshitz6}.  This statement has not been, 
and perhaps can never
be, rigorously proven for all theories, but is strongly supported by
the ample physical intuition coming from our experience with
macroscopic systems.  As the result of its universal nature,
hydrodynamics implies very precise constraints on the forms of the
correlation functions (in Minkowski space) of conserved currents and
components of the stress-energy tensor: basically, these correlators
are fixed once a few transport coefficients are 
known~\cite{KadanoffMartin,Yaffe}.

The reliance on Minkowski correlators, as opposed to Euclidean
(or Matsubara) Green's functions, is what separates this work from most of
previous work on finite-temperature AdS/CFT
correspondence.
The benefit of looking at
Minkowskian correlators is that while at finite temperature the
Euclidean correlation functions decay exponentially at large
separations, the Minkowski counterparts possess long-time
non-exponential tails which have universal character.

We shall perform a check that in the finite-temperature $\N=4$ SYM
theory, the hydrodynamic forms of the two-point Minkowskian
correlators are correctly reproduced by gravity.  This paper
deals with the diffusive modes only (for their definition see
section~\ref{sec:hydro}), which include the diffusion of R-charges and
the shear mode corresponding to transverse velocity fluctuations.  The
treatment of longitudinal velocity fluctuations, or sound waves, is
technically more complicated and is deferred to future work.

The paper is constructed as follows.  In section~\ref{sec:hydro} we
briefly review hydrodynamics equations and their implications for
thermal Green's functions.  In section~\ref{sec:grav} we describe the
gravity solution, and in section~\ref{sec:prescription} we review the
Minkowski prescription used to compute the Green's functions from gravity. 
In section~\ref{sec:Rcharge} we compute the correlators of
R-charge currents and show the emergence of the pole corresponding to
R-charge diffusion.  As a by-product we find the R-charge diffusion
rate.  In section~\ref{sec:shear} we discuss the hydrodynamic shear
mode; as in the case with the R-charge currents, this mode also has a
diffusive pole.  Section~\ref{sec:concl} contains concluding remarks.

\section{Hydrodynamic preliminaries}
\label{sec:hydro}

To see the constraints imposed by hydrodynamics on thermal correlation
functions, let us consider some theory, at finite temperature, with a
conserved global charge.  We denote the corresponding current as
$j^\mu$, so $j^0$ is the spatial density of the charge.  We assume
zero chemical potential for this charge, so in thermal equilibrium
$\<j^0\>=0$.  The quantity of interest is the retarded thermal Green's
function
\begin{equation}\label{GRcurrent}
  G^R_{\mu\nu} (\omega, \q) = -i\!\int\!d^4x\,e^{-iq\cdot x}\,
  \theta(t) \< [j_\mu(x),\, j_\nu(0)] \>\,,
\end{equation}
where $q=(\omega,\q)$, $x=(t,\x)$.
This function determines the response of the system on a small
external source coupled to the current.  When $\omega$ and $\q$ are
small, the external perturbation varies slowly in space and time, and
a macroscopic hydrodynamic description for its evolution is possible.
For example, $j^0$ evolves according to the diffusion equation (Fick's
law~\cite{Fick})
\begin{equation}\label{diffusion}
  \d_0 j^0 = D \nabla^2 j^0\,,
\end{equation}
where $D$ is a diffusion constant with dimension of length.  This
equation corresponds to an overdamped mode, whose dispersion relation
is
\begin{equation}\label{Rcharge-pole}
  \omega = - i D q^2 \,,
\end{equation}
which implies that there has to be a pole, located at $\omega=-iDq^2$
in the complex $\omega$-plane, in the retarded correlation functions
of $j^0$~\cite{KadanoffMartin}.

As a slightly more complicated example, let us consider the
correlators of the components of the stress-energy tensor
$T^{\mu\nu}$.  The linearized hydrodynamic equations have the form
\begin{equation}\label{hydro}
\begin{split}
  &  \d_0 \tilde T^{00} + \d_i T^{0i} = 0\,,\\
  &  \d_0 T^{0i} + \d_j \tilde T^{ij} = 0\,,
\end{split}
\end{equation}
where
\begin{equation}
\begin{split}
  \tilde T^{00} &= T^{00} - \epsilon, \qquad \epsilon =\< T^{00}\>\,,\\
  \tilde T^{ij} &= T^{ij} - P\delta^{ij}
   =- \frac1{\epsilon + P}\Bigl[\eta \Bigl(\d_i T^{0j} + \d_j T^{0i}
    -\frac23
    \delta^{ij}\d_k T^{0k}\Bigl) 
+ \zeta \delta^{ij}\d_k T^{0k}\Bigr],
\end{split}
\end{equation}
$\epsilon$ and $P$ are the energy density and pressure,
$\eta$ and $\zeta$ are the shear and bulk viscosities,
respectively.  eqs.~(\ref{hydro}) possess two types of eigenmodes: the
shear modes, which are the transverse fluctuations of the momentum
density $T^{0i}$, with a purely imaginary eigenvalue
\begin{equation}\label{shear-pole}
  \omega = - \frac{i\eta}{\epsilon+P} q^2 \,,
\end{equation}
and the sound wave, which is the simultaneous fluctuation of the
energy density $T^{00}$ and the longitudinal component of the momentum
density $T^{0i}$, with the dispersion relation
\begin{equation}\label{sound-pole}
  \omega = u_s q - \frac i2 \frac1{\epsilon+P} 
  \left(\zeta+\frac43\eta\right)q^2 \,,
  \qquad u_s^2 = \frac{\d P}{\d\epsilon}\, .
\end{equation}
The poles~(\ref{Rcharge-pole}) and~(\ref{shear-pole}) will be
reproduced from gravity calculations, while~(\ref{sound-pole}) is deferred
to future work.  If the theory is conformal, then the
stress-energy tensor is traceless, so $\epsilon=3P$ and $\zeta=0$.


\section{Gravity preliminaries}
\label{sec:grav}
The non-extremal three-brane background is a solution of the type 
IIB low energy equations of motion,
\begin{subequations}
\begin{eqnarray}
R_{MN} &=& {1\over 96} F_{MPQRS} F_N^{PQRS}\,,
  \label{a1}\\
 F_{(5)} &=& *  F_{(5)}\,,
  \label{a2}
\end{eqnarray}
\end{subequations}
where all other supergravity fields are consistently set to zero.
We use notations $M,N,...$ for the ten-dimensional indices, 
$\mu, \nu,...$ for the five-dimensional and $i,j,...$ 
for the four-dimensional ones. 
The solution is given by the  metric
\begin{equation}\label{non_extremal_metric}
  ds^2_{10} = H^{-1/2}(r)\left[ -f dt^2 + dx^2 + dy^2 +dz^2 \right]
+
   H^{1/2}(r) \left( f^{-1}dr^2 +r^2 d\Omega_5^2\right)\,,
\end{equation}
where $H(r) = 1+R^4/r^4$, $f(r) = 1- r_0^4/r^4$, and the Ramond-Ramond
 five-form,
\begin{equation}
F_5 = - {4 R^2\over H^2 r^5}(R^4+r_0^4)^{1/2} (1 + * )\,
 d t\wedge d x \wedge d y
 \wedge d z  \wedge d r\,,
\label{non_extremal_5form}
\end{equation}
with all other fields vanishing.

In the near-horizon limit $r\ll R$ the metric becomes
\begin{equation}\label{near_horizon_metric}
ds^2_{10} = 
  \frac{(\pi T R)^2}u
\left( -f(u) dt^2 + dx^2 + dy^2 +dz^2 \right) 
 +{R^2\over 4 u^2 f(u)} du^2 + R^2 d\Omega_5^2\,,
\end{equation}
where $T = r_0/\pi R^2$ is the Hawking temperature, 
and we have introduced $u = r_0^2/r^2$ and $f(u)=1-u^2$. 
The horizon corresponds to $u=1$,
the spatial infinity to $u=0$.

In the language of the gauge theory/gravity correspondence, the background
(\ref{near_horizon_metric}) with the non-extremality parameter $r_0$
is dual to the  ${\cal N}=4$ $SU(N)$ 
SYM at finite temperature  $T = r_0/\pi R^2$
in the limit of $N\rightarrow \infty$, $g^2_{YM}N \rightarrow \infty$.

\section{Prescription for Minkowskian correlators}
\label{sec:prescription}

From the previous discussion, one notes that in order to see the
emergence of hydrodynamic behavior, one needs to compute thermal Green's
functions in {\em Minkowski} space.  In ref.~\cite{gamma_paper} we
formulate and discuss in detail a prescription for computing two-point
Green's functions from gravity.  For convenience this prescription is
given here; technical details can be found in ref.~\cite{gamma_paper}.

First let us recall that in Euclidean space, the gravity/gauge theory
duality is encoded in the following equality,
\begin{equation}\label{orig-prescr}
\langle e^{ \int_{\partial M} \phi_0 \O}\rangle =
  e^{-S_{\rm cl}[\phi_0] }\,,
\end{equation}
where $\O$ is some boundary CFT operator and $\phi$ is the bulk field which
couples to it.

We do not claim to have a full Minkowskian analog of
eq.~(\ref{orig-prescr}).  Rather, we shall concentrate on two-point
functions, and formulate our prescription specifically for those
functions.  In Euclidean space, eq.~(\ref{orig-prescr}) implies that
finding $\<\O(x)\O(0)\>$ amounts to computing the second functional
derivative of $S_{\rm cl}$ on the boundary value $\phi_0$.  It can be
shown that the computation of $\<\O\O\>$ is reduced to the following
three steps.  We shall denote the radial coordinate as $u$, and for
definiteness assume that the boundary is located at $u=0$, and the
horizon is at some positive $u$.
\begin{itemize}
\item[i)] From the classical action for $\phi$ one extracts the function
$A(u)$ staying in front of $(\d_u\phi)^2$ in the kinetic term,
\begin{equation}\label{SclA}
  S_{\rm cl} = \frac12 \int\!du\,d^4x\,A(u)(\d_u\phi)^2+\cdots
\end{equation}
\item[ii)] Solving the linearized field equation for $\phi$, one
expresses the bulk field $\phi$ via its 
value of $\phi_0$ at the boundary,
\begin{equation}
  \phi(u,q) = f_q(u) \phi_0(q) \,,
\end{equation}
where we work in momentum space.  By definition, the mode function $f_q(u)$
is equal to 1 at $u=0$.
\item[iii)] The Euclidean Green's function is then
\begin{equation}\label{GE-prescr}
  G_E(q) = - A(u)f_{-q}(u)\d_u f_q(u) |_{u\to0}\,.
\end{equation}
\end{itemize}
One can see how these three steps work by recalling that the classical
action (\ref{SclA}), for classical solutions, reduces to the boundary
term $\sim A\phi\phi'$.  Note that in taking the limit $u\to0$ in
eq.~(\ref{GE-prescr}) one may need to throw away the contact terms.

Now we can proceed with the formulation of the prescription for
Minkowskian correlators.
\begin{itemize}
\item[i)] The same as in the Euclidean case.
\item[ii)] In Minkowski space one has to specify the boundary
condition at the horizon in addition to that at the boundary $u=0$.
We impose the incoming-wave boundary condition (waves are only
absorbed by the black branes but not emitted from there) for all
Fourier components $\phi_q$ with timelike $q$.  For spacelike $q$'s,
we require regularity at the horizon.

\item[iii)] The retarded thermal Green's function is
\begin{equation}\label{GR-prescr}
  G^R(q) = A(u)f_{-q}(u)\d_u f_q(u) |_{u\to0}\,.
\end{equation}
\end{itemize}
Choosing the outgoing-wave condition at the horizon would yield the
advanced Green's function $G^A$ instead.
The sign in eq.~(\ref{GR-prescr}) corresponds to the standard
convention of the retarded and advanced Green's functions,
\begin{equation}
\begin{split}
  G^R(\omega,\q) &= -i\!\int\!d^4x\, e^{-iq\cdot x}\,
  \theta(t) \<[\O(x),\, \O(0)]\> \,,\\
  G^A(\omega,\q) &= i\!\int\!d^4x\, e^{-iq\cdot x}\,
  \theta(-t) \<[\O(x),\, \O(0)]\>\,.
\end{split}
\end{equation}

In ref.~\cite{gamma_paper} we verify that the three steps outlined
above indeed give the correct retarded Green's functions in several
cases where independent verification is possible.  Admittedly, this
three-step prescription is aesthetically unsatisfactory: it cannot be
formulated as succinctly as eq.~(\ref{orig-prescr}).  Nevertheless,
it does seem to work.  One can hope that our prescription can be
embedded in future general framework which allows the calculation of
higher-point Green's functions as well.  Despite the shortcomings, the
prescription at hand is sufficient for the purpose of this paper.

\section{Thermal R-current correlators in ${\cal N}=4$ SYM
and R-charge diffusion}
\label{sec:Rcharge}

To compute the
current correlators, we use an approach similar to 
the one taken at zero temperature~\cite{Freedman:1998tz,Chalmers:1998xr},
the only difference being that 
we work in Minkowski rather than in the 
Euclidean space and use the {\it non}-extremal
supergravity background. 
Our starting point is the five-dimensional Maxwell action 
in the background  (\ref{near_horizon_metric}),
\begin{equation}
S = - \frac{1}{4 g_{SG}^2} \int d^5x \sqrt{-g} \, F_{\mu\nu}^a F^{\mu\nu\;a}\,,
\label{maxwell_action}
\end{equation}
where $g_{SG}=4\pi/N$ is the effective coupling constant fixed in 
\cite{Freedman:1998tz}. Reinstating powers of $R$, we have 
$g_{SG}^2 = 16 \pi^2 R/N^2$. 

We shall work in the gauge $A_u=0$.  After imposing this gauge
condition, one still can make the (residual) gauge transformations 
$A_\mu\to A_\mu
+\partial_\mu \Lambda$, where $\Lambda$ is independent of $u$.
We use the Fourier decomposition
\begin{equation}
A_{i} = \int\! {d^4 q\over (2\pi )^4} e^{-i\omega t + i \q\cdot\x}
A_{i}(q,u)\,.
\label{fourier}
\end{equation}
To simplify calculations, we also choose $\q$ along the $z$-direction
on the brane, so the four-momentum is $q = (\omega, 0,0,q)$.
Defining the dimensionless energy and momentum,
\begin{equation}
  \wn = \frac\omega{2\pi T}\,, \qquad \qn = \frac q{2\pi T} \,,
\end{equation}
the five-dimensional Maxwell equations,
\begin{equation}
\frac{1}{\sqrt{-g}} \partial_\nu [ \sqrt{-g} g^{\mu\rho} g^{\nu\sigma}
(\partial_\rho A_\sigma - \partial_\sigma A_\rho) ] = 0\,,
\end{equation}
reduce to the following set of the ordinary differential equations
\begin{subequations}
\begin{eqnarray}
  &&  \wn A_t' + \qn f \, A_z' = 0 \stru \,,\label{eq1}\\
  &&  A_t'' - \frac1{uf}\left( \qn^2 A_t + \wn \qn A_z \right) =0\,,
  \stru \label{eq2}\\
  &&  A_z'' + {f'\over f} A_z'  + \frac1{uf^2}
 \left(\wn^2 A_z + \wn \qn A_t \right) = 0\,,
  \stru \label{eq3}\\
  &&  A_{\alpha}'' + {f'\over f}  A_{\alpha}' + \frac1{uf}
 \left( \frac{\wn^2}{f} - \qn^2 \right) A_{\alpha} = 0\,, 
\label{eq4}
\end{eqnarray}
\end{subequations}
where $\alpha$ stands for either $x$ or $y$, and the prime denotes the
derivative with respect to $u$.  Not all these equations are
independent: eqs.~(\ref{eq1}) and (\ref{eq2}) imply
(\ref{eq3}). One can also check that they are invariant under residual
gauge transformations $A_t\to A_t-\omega\Lambda$, $A_z\to A_z+q\Lambda$.

One sees that $A_t$ and $A_z$ satisfy a coupled system of equations,
while each component of $A_{\alpha}$ satisfies a stand-alone equation.  The
R-charge diffusion appears only in the sector of $A_t$ and $A_z$.
One
can reduce eqs.~(\ref{eq1})--(\ref{eq3}) to a single equation by
expressing, from eq.~(\ref{eq2}), $A_z$ in terms of $A_t$,
\begin{equation}
A_z =  \frac{u f}{\wn \qn} A_t''-
 \frac\qn\wn A_t \, ,
\label{eq5}
\end{equation}
and then substituting $A_z$ into  eq.~(\ref{eq1}).  One obtains
\begin{equation}
  A_t''' + \frac{(uf)'}{uf} A_t'' + 
  {\wn^2 - \qn^2 f(u)\over  u f^2} A_t'=0\,,
\label{eq6}
\end{equation}
Thus eqs.~(\ref{eq4}), (\ref{eq5}) and  (\ref{eq6})
together with the boundary conditions determine 
all components of $A_i$ (up to a residual gauge transformation).

Equation (\ref{eq6}) is a second-order differential equation for $A_t'$ in
which $u=1$ is a singular point.  According to the general technique
for solving such equations, one first has to determine 
the singular
behavior of $A_t$.  If one substitutes into eq.~(\ref{eq6}) 
$A_t' = (1-u)^\nu F(u)$, where $F(u)$
is a regular function, one finds that only two values of $\alpha$ are
allowed: $\nu_{\pm}^{(1)}=\pm i\wn/2$.
The ``incoming wave'' boundary condition
at the horizon singles out $\nu_-$. The equation for $F(u)$ has the form
\begin{equation}\label{eq_for_F}
 F'' + \left( \frac{1-3u^2}{uf}+\frac{i \wn}{1-u}\right) F' 
+  \frac{i\wn (1+2u)}{2u f} F 
+
{\wn^2 [ 4 - u(1+u)^2]\over 4 u f^2}F
- {\qn^2\over u f}F = 0\,.
\end{equation}
In the long-wavelength, low-frequency limit, $\wn$ and $\qn$ are small 
parameters,
and the solution to eq.~(\ref{eq_for_F}) can be obtained perturbatively
as a double series in $\wn$ and $\qn^2$ (cf.~\cite{absorption}),
\begin{equation}\label{expansion}
F(u) = F_0 + \wn F_1 + \qn^2 G_1 + 
  \wn^2 F_2 + \wn \qn^2 H_1 + \qn^4 G_2 + \cdots\,.
\end{equation}
The explicit form for the first three terms, essential for obtaining
the leading-order expression for the correlators, is rather simple,
\begin{equation}
   F_0 = C\,,\quad
   F_1 = \frac{iC}2\ln\frac{2u^2}{1+u}\,,\quad
   G_1 = C\ln\frac{1+u}{2u}\,.
\end{equation}
The functions $F_2$, $G_2$ and $H_1$ are written explicitly 
in appendix~\ref{appendix1}. 

We now substitute our solution for  $A_t'$ into  eq.~(\ref{eq5})
and take the limit $u\rightarrow 0$ assuming the boundary conditions
\begin{equation}
\lim_{u\rightarrow 0} A_t(u) = A_t^0\,, \qquad 
\lim_{u\rightarrow 0} A_z(u) = A_z^0\,.
\end{equation}
This determines the constant $C$ in terms of $A_t^0$, $A_z^0$:
\begin{equation}
C =  \frac{\qn^2 A_t^0 + \wn \qn A_z^0}{Q(\wn,\qn)} \,,
\end{equation}
where $Q(\wn,\qn)$ has the following expansion over the small arguments,
\begin{equation}
  Q(\wn,\qn) = 
i\wn-\qn^2 + O(\wn^2, \wn\qn^2, \qn^4)\,.
\end{equation}
$C$ is obviously invariant under the residual gauge transformations.
Having thus found $A_t'(u)$,  we can obtain 
$A_z'(u)$ using eq.~(\ref{eq1}).

Similarly, the solution of eq.~(\ref{eq4}) is found to be
\begin{equation}
\begin{split}
   A_{\alpha} &= {8\, A_a^0 (1-u)^{-i\wn/2}\over 8 -2 i\wn \ln{2}
  + \pi^2 \qn^2}
 \biggl[ 1 +  \frac{i\wn}2 \ln\frac{1+u}2
 \\   &
\quad+ 
 {\qn^2\over 2} \Bigl( \frac{\pi^2}{12} + \mbox{Li}_2 (-u) 
  +
  \ln u \ln (1+u)
 + \mbox{Li}_2 (1-u)\Bigr) \biggr] 
+ O \left(\wn^2,\qn^4,\wn \qn^2\right)\,.
\end{split}
\end{equation}

Near the horizon, where $u=\epsilon$ is small, the solutions we found
imply the following relation between the radial derivatives of the
fields and their boundary values,
\begin{subequations}
\begin{eqnarray}
    A_\alpha' &=& i\omega A_\alpha^0 \,,\stru \\
    A_t' &=& (\qn^2A_t^0 + \wn\qn A_z^0)\ln\epsilon + 
    \frac{\qn^2A_t^0 + \wn\qn A_z^0}{i\wn-\qn^2}\,,\stru \\
    A_z' &=& -(\wn\qn A_t^0 + \wn^2 A_z^0)\ln\epsilon -
    \frac{\wn\qn A_t^0 + \wn^2 A_z^0}{i\wn-\qn^2}\,.
\end{eqnarray}
\end{subequations}
On the other hand, the terms in the action which contain two
derivatives with respect to $u$ are
\begin{equation}
\begin{split}
S &= - \frac{N^2}{32 \pi^2 R} 
\int\! du\,d^4x\, \sqrt{-g}\, g^{uu} g^{ij}  \d_u A_i \partial_u A_j
 +\cdots \stru
  \\
 &=  \frac{N^2 T^2}{16}\!
\int\! du\, d^4x\, [ A_t^{\prime 2} - 
f(A_x^{\prime 2} +  A_y^{\prime 2} + A_z^{\prime 2})]+\cdots
\end{split}
\end{equation}
Applying the prescription formulated in
Section~\ref{sec:prescription}, one finds
\begin{subequations}
\begin{eqnarray}
G_{x x}^{a b} &=& G_{y y}^{a b}  =- {i N^2 T\omega\; \delta^{a b}
\over 16\pi}
 + \cdots\stru
\,,\\
G_{t t}^{a b} &=&  { N^2 T q^2\; \delta^{a b}\over 16 \pi  ( i\omega - D q^2)}
 + \cdots\stru
\,,\\
G_{t z}^{a b} &=& G_{z t}^{a b}  = - 
  { N^2 T \omega q\; \delta^{a b}\over 16 \pi  ( i\omega - D q^2)}
 + \cdots\stru
\,,\\
G_{z z}^{a b} &=&  { N^2 T \omega^2\; 
  \delta^{a b}\over 16 \pi  ( i\omega - D q^2)}
 + \cdots
\,,
\end{eqnarray}
\end{subequations}
where $\cdots$ denotes corrections of order $\wn^2$, $\wn\qn^2$ or $\qn^4$,
and
\begin{equation}\label{D}
  D = \frac 1{2\pi T} \,.
\end{equation}

We see that the correlation functions of $j^0$ and $j^z$ contains the
diffusion pole expected from hydrodynamic arguments.  In contrast,
those of $j^x$ and $j^y$ do not have this pole.  The constant $D$ is
the diffusion constant of R-charges; its value in the large 
't Hooft coupling limit $g^2_{YM}N\rightarrow \infty$
is found explicitly in
eq.~(\ref{D}). This result can be regarded as  a nontrivial prediction 
for the strongly coupled ${\cal N}=4$ SYM at finite temperature. 
One observes that in this limit $D$ is independent of the 
't Hooft coupling, sharing this property with the shear 
viscosity (\cite{viscosity}, see also below).  Also, the diffusion constant
does not contain a power of $N$.
It is amazing that it is at all possible 
to compute a kinetic coefficient in a strongly coupled theory, and the
final result is as simple as eq.~(\ref{D}).  
What is also interesting
to note
is that the calculation of $D$ above is, arguably, technically simpler 
than similar calculations in weakly-coupled field
theories~\cite{Jeon,JeonYaffe,AMY,WangHeinz}.

It is interesting to compare eq.~(\ref{D}) with the result at weak
coupling.  In this regime, the diffusion constant is proportional to
the mean free path, which implies
\begin{equation}\label{D-weak}
  D \sim \frac1{(\gYM^2N)^2T\ln{1\over(\gYM^2N)}}\,,\qquad
  \gYM^2N\ll1\,.
\end{equation}
Eqs.~(\ref{D}) and (\ref{D-weak}) suggest that the behavior of $D$ as
a function of the 't Hooft coupling $\gYM^2N$ is
\begin{equation}\label{fD}
  D = f_D(\gYM^2N)\frac1T\,,
\end{equation}
where $f_D(x)\sim x^{-2}\ln^{-1}(1/x)$ for $x\ll1$ and
$f_D(x)=1/(2\pi)$ for $x\gg1$.

With some extra effort, one can also compute corrections to these
results in the parameters $\wn$ and $\qn$.  To the next order in
perturbation theory, the correlators which have the pole are given by
\begin{subequations}
\begin{eqnarray}
  G_{t t}^{a b} &=&  { N^2 T^2 \qn^2 P (\wn,\qn)
 \; \delta^{a b} \over 8\,  Q (\wn,\qn)}
\,, \stru \\
  G_{t z}^{a b} &=& G_{z t}^{a b} = - { N^2 T^2\wn\qn P(\wn,\qn)
 \; \delta^{a b} \over 8\,  Q (\wn,\qn)}
\,, \stru \\
  G_{z z}^{a b} &=&  { N^2 T^2 \wn^2 P (\wn,\qn)
 \; \delta^{a b} \over 8\,  Q (\wn,\qn)}
 \,,
\end{eqnarray}
\end{subequations}
where corrections of order 
${\cal O}(\wn^3,\wn^2\qn^2, \wn\qn^4, \qn^6)$ and higher are omitted, and
\begin{equation}
\begin{split}
P (\wn,\qn) &= 1+ \ln2\, \Bigl(\frac i2\wn -\qn^2\Bigr)
\,,\\
Q(\wn,\qn) &= i\wn - \qn^2 +\ln2\,\Bigl(\frac{\wn^2}2
  +\frac i2 \wn\qn^2 - \qn^4\Bigr)\,.
\end{split}
\end{equation}
The zeroes of
$Q(\wn,\qn)$ determine the 
poles of the correlators.  For small $\qn$ only 
one of the zeros has a value compatible 
with our assumption $\wn \ll 1$, others must be discarded. The position of
the pole is given by
\begin{equation}
  \wn = - i\qn^2(1+\qn^2\ln2)\,,
\end{equation}
or, in terms of the dimensionful energy and momentum $\omega$, $q$,
\begin{equation}
i\omega = D q^2 \left( 1 + {q^2 \ln{2}\over 4 \pi^2 T^2}\right)\,.
\end{equation}

\section{Near-extremal metric perturbations and the shear mode}
\label{sec:shear}

To compute the two-point function of the stress-energy tensor in 
the boundary theory, we consider a small 
perturbation of the five-dimensional part of the near-extremal background 
(\ref{near_horizon_metric}), $g_{\mu\nu} = 
g_{\mu\nu}^{(0)}+ h_{\mu\nu}$, where  $g_{\mu\nu}^{(0)}$ is given by
\begin{equation}
ds^2_{5} = {\pi^2 T^2 R^2\over u}
\left( -f(u) dt^2 
  + d\x^2 \right) +
 {R^2\over 4 f(u) u^2} du^2\,.
\label{5dbg}
\end{equation}
The part of the ten-dimensional metric corresponding to $\mathrm{S}^5$ remains 
unperturbed, and the 
five-form field (\ref{non_extremal_5form}) is unchanged to the first order in 
perturbation. The perturbed metric satisfies
\begin{equation}
{\cal R}_{\mu\nu} = {\cal R}_{\mu\nu}^{(0)} + {\cal R}_{\mu\nu}^{(1)}+\cdots
= {2\Lambda\over 3} g_{\mu\nu}\,,
\end{equation}
where  $\Lambda = -6/R^2$. To the first order in  $h_{\mu\nu}$ the 
Einstein equations are
\begin{equation}
 {\cal R}_{\mu\nu}^{(1)} = - {4\over R^2} h_{\mu\nu}\,.
\label{first_order_es}
\end{equation}
The solution of the Dirichlet problem for 
eq.~(\ref{first_order_es}) is then
substituted into the $5d$ gravitational action expanded near the background 
(\ref{5dbg}). The action is given by
\begin{equation}\label{grav_action}
S  = {\pi^3 R^5\over 2 \kappa^2_{10}}\Biggl[ \int\limits_{0}^{1}\!du\! \int\! 
d^4x\, \sqrt{-g}
\left( {\cal R} {-} 2\Lambda\right)
 + 
 2\!\int\! d^4x\, \sqrt{-h}\, K \Biggr].
\end{equation}
Here $\kappa_{10} = \sqrt{8\pi G}$ is the ten-dimensional
 gravitational constant,
related to the parameter $R$ of the non-extremal geometry and the number
$N$ of coincident branes by $\kappa_{10} = 2\pi^2\sqrt{\pi}R^4/ N$
\cite{Gubser:1996de}.
  The second integral is the Gibbons-Hawking boundary term
with $K$ being the trace of the extrinsic curvature of the boundary. 
%
%

Again, we shall assume the perturbation to be dependent only on $t$
and $z$ (and proportional to $e^{-i\omega t+iqz}$), and choose the gauge
where $h_{u\mu}=0$ for all $\mu$.  In this case, the
gravitational perturbation can be classified by the spin under the
O(2) rotations in the $xy$ plane.  Specifically, there are three classes
of perturbations (only nonzero components of $h_{\mu\nu}$ are listed):
\begin{itemize}
\item $h_{xy}\neq0$, or $h_{xx}=-h_{yy}\neq0$;
\item $h_{xt}$ and $h_{xz}\neq0$, or $h_{yt}$ and $h_{yz}\neq0$;
\item $h_{tz}$ and all diagonal elements of $h_{\mu\nu}$ are nonzero, 
and $h_{xx}=h_{yy}$.
\end{itemize}
The field equations in each of the classes decouple from the other ones.
We shall consider the first two cases.  The third case is related to the
sound wave in field theory and is not considered in this paper.

\subsection{Off-diagonal perturbation with $h_{xy}\neq 0$}

The simplest case we consider is that when
the off-diagonal perturbation  $h_{xy}\neq 0$, and all other perturbations
vanish. The equations for the perturbation with $h_{xx}=-h_{yy}$ is
absolutely identical.
eq.~(\ref{first_order_es}) becomes
\begin{equation}
h_{xy}'' + {1-3u^2\over u f} h_{xy}' + {1\over (2\pi T)^2f^2 u}
 \biggl(f{\partial^2 
h_{xy}\over \partial z^2} - 
{\partial^2 
h_{xy}\over \partial t^2}\biggr) 
- {1+u^2\over f u^2}h_{xy} = 0\label{xy_perturbation}\,.
\end{equation}
Introducing a new function $\phi = u h_{xy}/(\pi T R)^2$ 
(i.e., $\phi = h^x_y$), and using 
the Fourier component $\phi_k(u)$ defined as in 
(\ref{fourier}), we observe that
eq.~(\ref{xy_perturbation})
becomes an equation for a minimally coupled scalar in the background
(\ref{5dbg}):
\begin{equation}
\phi_k'' - {1+u^2\over uf}\phi_k' 
+ {\wn^2 -\qn^2f\over  u f^2}
\phi_k = 0\,.
\end{equation}
Therefore, the computation now is completely similar to the
computation of the Chern-Simon diffusion rate in ref.~\cite{gamma_paper}.
The solution representing the incoming wave at the horizon is given by
\begin{equation}
\phi_k(u) = (1-u)^{-i\wn /2} F_k(u)\,,
\end{equation}
where $F_k(u)$  is regular at $u=1$ and 
can be written as a series \cite{gamma_paper}
\begin{equation}
F_k(u) = 1 - {i\wn \over 2} \ln { {1+u\over 2}} 
+ 
{\wn^2\over 8} \Biggl[ \left( \ln{ {1+u\over 2 }} + 8 (1-{\qn^2\over \wn^2}
)\right)
 \ln{ {1+u\over 2 }}
- 4\, \mbox{Li}_2\, {1-u\over 2} \Biggr] 
+ O(\wn^3)\,,
\end{equation}
where $\mbox{Li}_2(z)$ is the polylogarithm.

The term proportional to $\phi^{\prime 2}$ in the action is
\begin{equation}
S = - {\pi^2 N^2  T^4\over 8 }\! \int\!du\, d^4 x\, {f\over u} 
    \phi^{\prime 2} +\cdots
\end{equation}

Let us define, in complete analogy with eq.~(\ref{GRcurrent}), the
retarded Green's function for the components of the stress-energy tensor,
\begin{equation}
  G_{\mu\nu,\lambda\rho} (\omega, \q)
  = -i\!\int\!d^4x\,e^{-iq\cdot x}\,
  \theta(t) \< [T_{\mu\nu}(x),\, T_{\lambda\rho}(0)] \>\,.
\end{equation}
Then, according to the prescription given in~\cite{gamma_paper}, the
(retarded) two-point function 
of the stress-energy tensor (in Fourier space) reads
\begin{equation}
  G_{xy,xy}(\omega,\q) =
  - { N^2 T^2 \over 16 }\left( i\, 2 \pi T \omega + q^2 \right)  \,.
\end{equation}

This result can be used to
determine the shear viscosity of the strongly coupled 
${\cal N}=4$ SYM plasma. 
One has to recall the Kubo formula, which relates
the shear viscosity to
the correlation function of the stress-energy tensor at zero 
spatial momentum,
\begin{equation}\label{eta-hxy}
\eta = \lim_{\omega \rightarrow 0} {1\over 2\omega} 
 \int\!dt\,d\x\, e^{i\omega t}\,
\langle [ T_{xy}(x),\,  T_{xy}(0)]\rangle =  
  \frac{\pi}8 N^2 T^3\,.
\end{equation}
The result coincides with the one obtained in~\cite{viscosity}
from the absorption calculation, and also from the value obtained from
the pole of the propagator (see below).

\subsection{Off-diagonal perturbation with  $h_{tx}\neq 0$, $h_{xz}\neq 0$}

This is the channel where the correlation functions have the
diffusion pole.
In this case the Einstein equations (\ref{first_order_es}) give 
the following set of coupled equations for the Fourier components of the 
perturbations 
$H_t = u h_{t x}/(\pi T R)^2$ and $H_z = u h_{z x}/(\pi T R)^2$:
\begin{subequations}
\begin{eqnarray}
&&H_t' + {\qn f\over \wn} H_z' = 0 \,,
\label{grav_1} \stru \\
&&
H_t'' - {1\over u} H_t' - {\wn \qn \over u f} H_z - {\qn^2\over uf} H_t = 0\,,
\label{grav_2}\stru \\
&&H_z'' - {1+u^2\over u f} H_z' + {\wn^2  \over u f^2} H_z +
 {\wn\qn\over uf^2} H_t = 0\,.
\label{grav_3}
\end{eqnarray}
\end{subequations}
This system is very similar to the one encountered in the 
previous Section for the potentials $A_t$, $A_z$. Eqs.~(\ref{grav_1}) and
(\ref{grav_2}) imply  (\ref{grav_3}).
Solving  (\ref{grav_2}) for $H_z$ we get
\begin{equation}
H_z = {u f\over \wn\qn} H_t'' - {f\over \wn\qn} H_t' -{\qn\over \wn} H_t\,.
\end{equation}
Substituting this into  eq.~(\ref{grav_1}) gives
\begin{equation}
H_t''' - {2 u\over f} H_t'' + {2 u f - \qn^2 f +\wn^2\over u f^2} H_t' =0\,.
\end{equation}
In the low-frequency, long-wavelength limit, this equation is solved by making 
the substitution $H_t'(u) = (1-u)^{-i\wn/2}G(u)$ 
and then solving perturbatively the resulting equation for $G(u)$, 
\begin{equation}
G'' - \left( {2u\over f} - {i\wn\over 1-u}\right) G'
 + \frac1f \biggl( 2 
+ {i \wn\over 2 } - {\qn^2\over u}
+ {\wn^2 [ 4 - u(1+u)^2]\over 4 u f} \biggr) G = 0\,.
\end{equation}
The solution regular at $u=1$ is
\begin{equation}
G(u) = C\left[ u - i\wn \biggl( 1-u-{u\over 2}\ln{ {1+u\over 2}}\biggr) + 
{\qn^2(1-u)\over 2}\right] 
+ O \left( \wn^2,\wn\qn^2, \qn^4 \right)\,.
\end{equation}
Taking the limit $u\rightarrow 0$ in the eq.~(\ref{grav_2}) we find 
 $C$ in terms of the boundary values $H_t^0$ and $H_z^0$:
\begin{equation}
C = {\qn^2 H_t^0 + \qn\wn H_z^0\over i\wn - {\qn^2\over 2}}\,.
\end{equation}

The terms proportional to $H_t^{\prime 2}$ and $H_z^{\prime 2}$ in the
action~(\ref{grav_action}) are
\begin{equation}
S = - {\pi^2 N^2  T^4\over 8 } \int\!du\, d^4 x\,\frac1u 
  \Bigl( - H_t^{\prime 2} +  f H_z^{\prime 2}\Bigr) +\cdots
\end{equation}
Again, as in the previous Section, all constant and contact terms are 
ignored.
Computing the correlators, we get
\begin{subequations}\label{corr-T}
\begin{eqnarray}
  G_{tx,tx}(\omega,\q) &=&  
  { N^2\pi T^3 q^2 \over 8 (i\omega - {\cal D} q^2)}
\,,
\stru\\
 G_{tx,xz}(\omega,\q)  &=& 
- { N^2\pi T^3 \omega q \over 8 (i\omega - {\cal D} q^2)}
\,,\stru\\
 G_{xz,xz}(\omega,\q) &=& 
 { N^2\pi T^3 \omega^2  \over 8 (i\omega - {\cal D} q^2)}
\,,
\end{eqnarray}
\end{subequations}
where the formulas are valid up to corrections of order 
${\cal O} (\wn^2, \wn\qn^2, \qn^4 )$, and
\begin{equation}\label{shear-D}
   {\cal D} = \frac1{4\pi T}\,,
\end{equation}
i.e., is twice smaller than the diffusion constant for the R-charge.

According to formula~(\ref{shear-pole}), ${\cal D}=\eta/(\epsilon+P)$.
The pressure $P$ can be found from the expression for the entropy
density~\cite{Gubser:1996de}, which differs from that at zero gauge
coupling by a well-known factor of $3/4$,
\begin{equation}\label{entropy}
  s = \frac34 s_0 = \frac{\pi^2}2 N^2 T^3\,,
\end{equation}
and the thermodynamic relation $s=\d P/\d T$.  Moreover,
$\epsilon=3P$, thus
$\epsilon+P=\frac{\pi^2}2N^2T^4$. Equations~(\ref{shear-pole}) and
(\ref{shear-D}) then yield $\eta=\frac\pi8N^2T^3$, which agrees
with ref.~\cite{viscosity} and eq.~(\ref{eta-hxy}).  Thus, we have
seen the consistency of several seemingly unrelated calculations: the
calculation of the entropy, the calculation of the viscosity via
absorption, the calculation of the
correlators~(\ref{corr-T}), and the determination of the pole from
linearized hydrodynamics.  In particular, if the entropy differed from
the free-gas value by a factor other than 3/4, such a consistency
would not have been seen.

Again, the comparison of our result for $\eta$ with that in the
weak-coupling regime~\cite{Jeon,JeonYaffe,AMY,WangHeinz} suggests that
the behavior of $\eta$ as a function of the 't Hooft coupling is
\begin{equation}\label{feta}
  \eta = f_\eta(\gYM^2N) N^2T^3
\end{equation}
where $f_\eta(x)\sim x^{-2}\ln^{-1}(1/x)$ for $x\ll1$ and
$f_\eta(x)=\frac\pi8$ for $x\gg1$.  Equations~(\ref{fD}) and
(\ref{feta}) suggest that the $\alpha'$ corrections to $D$ and $\eta$
are positive.

\section{Conclusion}\label{sec:concl}

In this paper, we have used a Minkowski AdS/CFT prescription to
compute the real-time Green's functions of the R-charge currents and the
components of the stress-energy tensor in the $\N=4$ SYM theory at
strong coupling and finite temperature.  We showed that the results
agree with the expectations from hydrodynamic theory, which can be
interpreted as a verification of the Minkowski prescription, or of the
existence of the hydrodynamic behavior in the finite-temperature field
theory, or both.  A more general conclusion is that our results support 
the validity of the finite-temperature AdS/CFT correspondence, even
in four dimensions
where no independent checks based on non-renormalization 
arguments were known to exist.
We hope that the calculations performed in this
paper can be extended to cover the sound wave, as well as to other
examples of gravity/gauge theory duality.

\begin{acknowledgments}
The authors thank I.R.~Klebanov, P.K.~Kovtun, M.~Porrati, and L.G.~Yaffe for
discussions, and C.P.~Herzog for his comments on the manuscript.
This work is supported, in part, by DOE Grant No.\
DOE-ER-41132.  The work of D.T.S.\ is supported, in part, by the Alfred P.\
Sloan Foundation. G.P. is supported in part by C.N.R.--Italy Grant. 
\end{acknowledgments}

\appendix

\section{Perturbative solution for $A_t'(u)$}
\label{appendix1}
Here we give explicit expressions for the first few terms 
of the perturbative expansion (\ref{expansion}). 
Integration constants are fixed by requiring  functions 
$F_1$, $F_2$, $G_1$, $G_2$, $H_1$
 to be regular
at the horizon ($u=1$), and to have a vanishing limit as $u\rightarrow 1$.
\begin{equation} \label{solution3}
\begin{split}
F_2 &= {C\over 24} \biggl[ \pi^2 + 3 \ln^2 2 + 3\ln^2 (1+u) 
 + 6\ln{2}\ln{{u^2\over 1+u}}
 \\
 &\quad 
+ 12\, \mbox{Li}_2(1-u) + 12\, \mbox{Li}_2(-u)- 
 12\, \mbox{Li}_2\,{1-u\over 2}\biggr]\,,
\end{split}
\end{equation}
\begin{equation}\label{solution4}
\begin{split}
H_1 &= i C \biggl[ -\frac{\pi^2}{12} - i \pi \ln{2} + \ln^2 2 
 + \frac12 \ln{(1+u)}\ln{u(1-u)^2\over 4}
\\
  &\quad 
  + \frac{\ln2}2\ln{u\over (u-1)^2}+
   \mbox{Li}_2(-u)+ 
  \mbox{Li}_2\left(1-u\right) + 
  \mbox{Li}_2\,{1+u\over 2} \biggr],
\end{split}
\end{equation}

\begin{equation} \label{solution5}
\begin{split}
G_2 &= C \biggl[ - \frac{\pi^2}{24} 
 + \frac12\ln^2u - \frac12\ln u \ln(1+u)
 -\ln\biggl(-\frac 1u\biggr)\ln[2u(1+u)]\\
 &\quad +  \mbox{Li}_2 (2)  
 - \mbox{Li}_2\left(1+{1\over u}\right) + 
 \, \mbox{Li}_2\, {1-u\over 2} 
 -\frac12[  \mbox{Li}_2 (1-u) + \mbox{Li}_2(-u)]
\biggr]\,. 
\end{split}
\end{equation}

\end{document}